\documentclass[journal]{IEEEtran}
\usepackage{amsmath,amsfonts}
\usepackage{algorithmic}
\usepackage{algorithm}
\usepackage{array}
\usepackage[caption=false,font=normalsize,labelfont=sf,textfont=sf]{subfig}
\usepackage{textcomp}
\usepackage{stfloats}
\usepackage{url}
\usepackage{verbatim}
\usepackage{graphicx}
\usepackage{cite}
\usepackage{acronym}
\usepackage{makecell}

\usepackage[hidelinks]{hyperref}

\usepackage{ifpdf}
\usepackage{url}
\ifpdf
\else
\fi

\usepackage{cite}
\usepackage{balance}
\usepackage{bm,comment,color}

\usepackage{soul}
\usepackage{amsmath}
\usepackage{array}
\usepackage{amsfonts} 
\usepackage{amssymb}
\usepackage{esint} 
\usepackage{units}
\usepackage{multirow}
\usepackage{comment}

\usepackage[table]{xcolor}
\usepackage{hhline}

\usepackage{color}

\newcommand{\yellow}[1]{{\color{yellow}{#1}}}

\usepackage{pgfplots}
\usepackage{tikz}
\usetikzlibrary{calc}
\makeatletter
\newcommand{\gettikzxy}[3]{%
  \tikz@scan@one@point\pgfutil@firstofone#1\relax
  \edef#2{\the\pgf@x}%
  \edef#3{\the\pgf@y}%
}
\usetikzlibrary{spy,backgrounds}

\pgfplotsset{compat=newest}
\usetikzlibrary{plotmarks}
\usetikzlibrary{arrows.meta}
\usepgfplotslibrary{patchplots}
\usepackage{grffile}
\newlength\fheight 
\newlength\fwidth 
\usepgfplotslibrary{fillbetween}

\acrodef{6g}[6G]{the sixth generation}
\acrodef{aoa}[AOA]{angle-of-arrival}
\acrodef{bs}[BS]{base station}
\acrodef{bse}[BSE]{beam squint effect}
\acrodef{elaa}[ELAA]{extremely large antenna array}
\acrodef{ff}[FF]{far-field}
\acrodef{las}[L\&S]{localization and sensing}
\acrodef{los}[LOS]{line-of-sight}
\acrodef{nf}[NF]{near-field}
\acrodef{nlos}[NLOS]{non-line-of-sight}
\acrodef{ofdm}[OFDM]{orthogonal frequency division multiplexing}
\acrodef{ris}[RIS]{reconfigurable intelligent surface}
\acrodef{sns}[SNS]{spatial non-stationarity}
\acrodef{swm}[SWM]{spherical wave model}
\acrodef{siso}[SISO]{single-input single-output}
\acrodef{ue}[UE]{user equipment}
\acrodef{dmimo}[D-MIMO]{distributed MIMO}
\long\def\comment#1{}

\newfont{\bbb}{msbm10 scaled 700}

\newcommand{\vthickline}{\vrule width 0.8pt}
\newcommand{\hthickline}{\noalign{\hrule height 0.80pt}}

\newfont{\bb}{msbm10 scaled 1100}

\newcommand{\pv}{{\bf p}}
\newcommand{\qv}{{\bf q}}

\makeatletter
\def\ps@IEEEtitlepagestyle{%
\def\@oddfoot{\mycopyrightnotice}%
\def\@evenfoot{}%
}
\def\mycopyrightnotice{%
{
\scriptsize {\copyright~2023 IEEE. Personal use of this material is permitted. Permission from IEEE must be obtained for all other uses. More details can be found in IEEE Post-publication policies.}
} 
\gdef\mycopyrightnotice{}
}

\makeatletter
\let\old@ps@headings\ps@headings
\let\old@ps@IEEEtitlepagestyle\ps@IEEEtitlepagestyle
\def\confheader#1{
\def\@oddhead{\strut\hfill#1\hfill\strut}%
}
\makeatother
\confheader{
\scriptsize \shortstack{This article has been accepted for publication in a future issue of IEEE Wireless Communications, but has not been fully edited. Content may change prior to final publication.}
}

\begin{document}

\title{6G Localization and Sensing in the  Near Field: Features, Opportunities, and Challenges}

\author{
Hui~Chen,~\IEEEmembership{Member,~IEEE},
Musa~Furkan~Keskin,~\IEEEmembership{Member,~IEEE}, 
Adham~Sakhnini,
Nicol\`o~Decarli,~\IEEEmembership{Member,~IEEE},
Sofie~Pollin,~\IEEEmembership{Senior~Member,~IEEE},
Davide~Dardari,~\IEEEmembership{Senior~Member,~IEEE},
and~Henk~Wymeersch,~\IEEEmembership{Fellow,~IEEE}

\thanks{This paper is partially supported by the  Vinnova B5GPOS Project under Grant 2022-01640, the European Union under the Italian National Recovery and Resilience Plan
(NRRP) of NextGenerationEU, partnership on “Telecommunications of the Future” (PE00000001 - program “RESTART”), and by the EU Horizon project TIMES (Grant no. 101096307).}
}

\maketitle

\begin{abstract}
The far-field channel model has historically been used in wireless communications due to the simplicity of mathematical modeling and convenience for algorithm design. With the need for high data rates, low latency, and ubiquitous connectivity in the sixth generation (6G) of communication systems, new technology enablers such as extremely large antenna arrays (ELAAs), reconfigurable intelligent surfaces (RISs), and distributed multiple-input-multiple-output (D-MIMO) systems will be adopted. These enablers not only aim to improve communication services but also have an impact on localization and sensing (L\&S), which are expected to be fundamentally built-in functionalities in future wireless systems. Despite appearing in different scenarios and supporting different frequency bands, such enablers share the so-called near-field (NF) features, which will provide extra geometric information conducive to L\&S. In this work, we describe the NF features, namely, the spherical wave model, spatial non-stationarity, and beam squint effect. After discussing how L\&S see NF differently from communication, the opportunities and open research challenges are provided.
\end{abstract}


\section{Introduction}
\Ac{6g} systems are expected to support immense throughput, ultra-massive communications with high connection density~\cite{zhang20236g}.
The wide bandwidths in the mmWave and sub-THz bands can be leveraged to meet the high data rates and low latency requirements for communications. In this context, \acp{elaa} can be used to combat the high propagation losses. Moreover, technologies such as~\acp{ris} and~\ac{dmimo} are also under research for omnipresent connectivity and extended coverage. 
In addition to their benefits from the communications perspective, these technologies also offer unprecedented opportunities for \ac{las}, constituting the core functions of 6G networks.

\Ac{las} are defined as the {state estimation (e.g., position, orientation, and velocity estimation)} of the connected devices (i.e., \textit{localization} of \acp{ue}) and passive {targets} (i.e.,  radar-like {monostatic or device-free bistatic/multistatic} \textit{sensing} of passive objects) {in a global frame of reference}.
In addition to the high delay resolution {(inverse of the bandwidth)} provided by the wideband signals, the large apertures of \acp{elaa} and \acp{ris} provide unparalleled angular resolution {(signal wavelength divided by the physical array aperture size)}, which directly improves the \ac{las} performance. Moreover, by acting as extra anchor nodes, \acp{ris} can support and enable various (and sometimes extremely challenging) scenarios, such as \ac{siso}~\cite{dardari2022nlos} and out-of-coverage systems~\cite{chen2023riss}. 
As for \ac{dmimo}, the spatial diversity and the exploitation of phase-coherence can significantly boost \ac{las} performance~\cite{nearfieldSense_TWC_2022}.
{The advent of these technologies increases the likelihood of encountering \ac{nf} scenarios, emphasizing the necessity for accurate NF models and algorithms to fulfill the requirements of communication, localization, and sensing in 6G systems.}

There are three features that make the NF scenario distinct, namely, \ac{swm}, \ac{sns}, and \ac{bse}. 
In particular, the validity of the plane wave approximation is violated when the communication distance is short and the antenna arrays are electrically large (i.e., with respect to the wavelength)~\cite{dardari2022nlos}. In this case, the SWM (or wavefront curvature) has to be taken into account.
In addition, due to \ac{nf} scattering, local fading, and blockages, different antenna elements observe the signal propagation from the same source differently, thus resulting in a \ac{sns} across the array~\cite{de2020non, tian2023low}.
Finally, when such arrays operate with wide bandwidths and frequency-independent phase shifters, the \ac{bse} causes different signal frequency components to be focused towards different areas~\cite{cui2022near}.

\begin{figure*}[t]
\centering
\centerline{\includegraphics[width=0.8\linewidth]{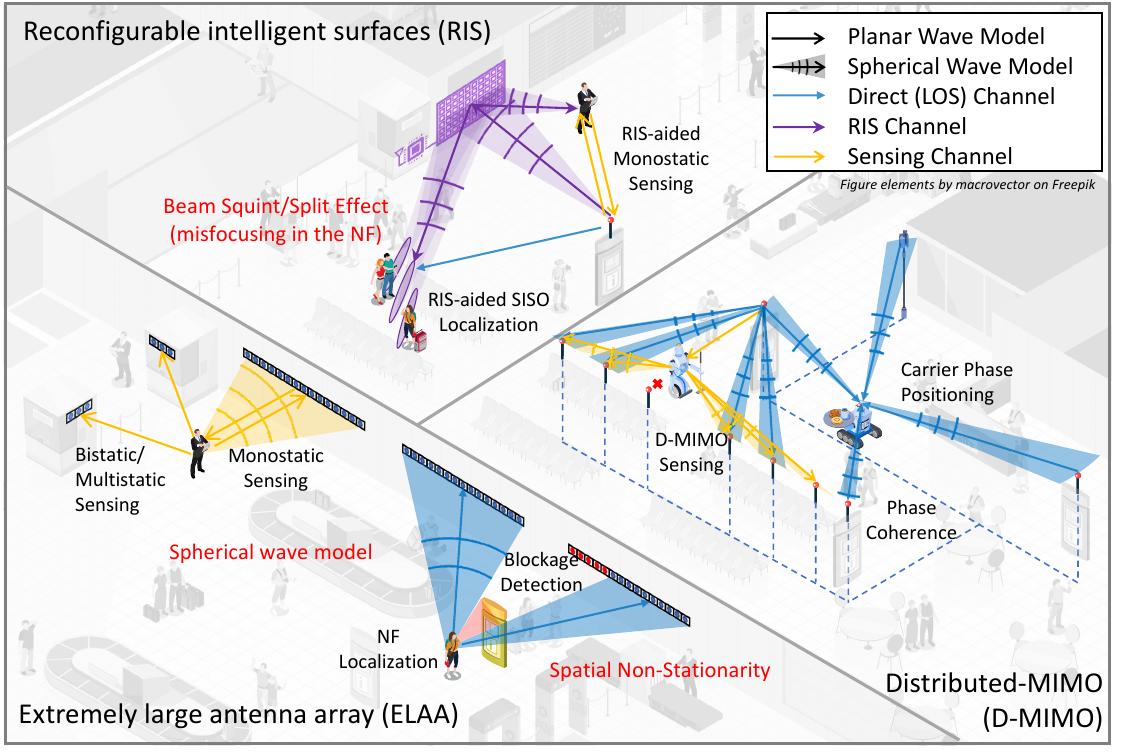}}
\caption{Illustration of NF \ac{las} in an indoor (airport terminal) scenario. \ac{las} services can be provided with ELAAs, RIS, and phase-coherent D-MIMO. In the NF, features such as SWM and SNS become prominent. With large bandwidths, the BSE manifests itself as misfocusing. Phase coherence makes the D-MIMO system function as a spatially distributed antenna array (a special form of ELAA). 
} \vspace{-5mm}
\label{fig-1-illustration}
\end{figure*}

When ignoring these features and adopting conventional \ac{ff} processing, model mismatch happens, causing significant performance degradation, especially when the SNR is high~\cite{chen2022channel}. On the other hand, the NF features, when correctly accounted for, can be harnessed to improve communication and L\&S performance.
For example, significant efforts have been invested in \ac{nf} communications, {as evidenced by studies highlighting its benefits in} mitigating co-channel interference and improving spectral efficiency~\cite{wang2023extremely, zhang20236g}. 
For \ac{las}, \ac{nf} features have also been exploited, e.g., \ac{swm} to support narrowband systems~\cite{nearfieldSense_TWC_2022} and \ac{los}-blocked scenarios~\cite{dardari2022nlos}, and \ac{sns} for blockage detection~\cite{tian2023low}. However, an overview of the \ac{nf} \ac{las} opportunities (e.g., reduced hardware and extended coverage) and challenges (e.g., high model complexity, calibration processes, and performance-cost trade-offs) is still lacking.

This work focuses on the discussion of \ac{nf} \ac{las} supported by \acp{elaa}, \acp{ris}, and phase-coherent~\ac{dmimo}, as shown in Fig.~\ref{fig-1-illustration}. Before delving into \ac{las}, we will briefly describe \ac{nf} features, and summarize the opportunities and challenges brought to communications. Then, from the \ac{las} perspective, we elaborate on the opportunities by listing several enabling scenarios in the NF with ELAA, RIS, and D-MIMO technologies. The open research challenges will also be discussed, followed by the conclusion of this work.

\section{Near-field Channel Features}
\label{sec_nf_model}
This section provides a recap of the three main features of \ac{nf} systems (i.e., the \ac{swm}, \ac{sns}, and \ac{bse}) and their effects on communications. Detailed discussions of the effect of these features on \ac{las} will be provided in the later sections.

\subsection{Spherical Wave Model}

One of the major distinguishing features of \ac{nf} propagation conditions is the spherical wavefront, 
characterized by the \ac{swm}. In standard \ac{ff} conditions (in general, beyond the Fraunhofer distance), phase variations across the receiving array depend only on the source angle with respect to the array.
In contrast, in the \ac{nf}, the phase across the array depends on the source location (i.e., coupled angle and distance information). 
This provides communications with multiple access capabilities where \acp{ue} at the same angle but different distances can be served~{without inter-UE interference} (i.e., beamfocusing)~\cite{zhang20236g}. In addition, multiple data streams can be exchanged in a single \ac{los} channel to improve the overall spectral efficiency. A small beamfocusing area can also mitigate interference and protect privacy at the expense of increased channel estimation complexity.

\subsection{Spatial Non-Stationarity}
The \ac{sns} occurs in the \ac{nf} because different regions of large arrays see different propagation paths~\cite{de2020non}.
Different from the \ac{ff} assumption, where the whole array receives signals from a specific path with the same amplitude and angle-dependent phase profile, the channel gains need to be characterized at the antenna level (or subarray level) instead of the array level to account for the \ac{sns}. 
{In the NF regime}, non-isotropic scattering of spatially extended targets {renders the SNS effect more prominent}~\cite{Fishler2006}. This is caused by angle-dependent {scattering} and self-occlusion of the individual scatterers that compose the target, {inducing} varying, often nearly random and highly fluctuating phases and amplitudes {across the array}.

In addition to the environment, imperfect manufacturing (i.e., non-isotropic radiation patterns) and element failures in large arrays also contribute to the \ac{sns}. 
The \ac{sns} does not adversely impact communications since the channel is viewed in an aggregated form by the transmitter-receiver couple. Indeed, the overall spectral efficiency of the system can be improved due to the rich propagation paths experienced by the large array.

\subsection{Beam Squint Effect}
\label{sec:BSE}
The \ac{bse} is caused by analog phase shifters that apply the same phase shift for signals at different frequencies. {When the signal bandwidth is large, the fixed antenna spacing induces frequency-dependent phase shifts across the array, causing the \ac{bse}.} In the \ac{ff}, the \ac{bse} results in the split beams for different subcarriers in different directions rather than the intended one. In contrast to the split beams in the \ac{ff}, the \ac{nf} scenario \ac{bse} makes beams at different frequencies {focusing on locations other than the intended one}~\cite{cui2022near}.
{For communications, the BSE can have a detrimental effect since the intended user is not properly served due to the energy leakage at certain frequencies {due to beam misfocusing}. This effect is particularly pronounced in the \ac{nf} when sharply focused beams are adopted. On the other hand, the BSE can be leveraged for faster initial access when establishing communication links {due to the wider coverage area by different subcarriers~\cite{cui2022rainbow}.}}

\subsection{Visualization of NF Features}
An illustration of NF channel features under different system parameters is shown in Fig.~\ref{fig-2-nf_features}. The \ac{swm} is reflected as focused points instead of angular beams as shown in Fig.~\ref{fig-2-nf_features} (b), and the effect of \ac{sns} caused by partial antenna blockage as shown in Fig.~\ref{fig-2-nf_features} (c). 
We can see from the figure that the SWM (resulting {in} focused points) is more pronounced in small angles and larger arrays. Additionally, the \ac{bse} is prominent at large angles, with edge subcarriers, and in large arrays (the squinted angles are the same, but the beamwidth in large arrays is smaller). Furthermore, the \ac{sns} caused by the partial blockage may affect the beamfocusing performance, but the communication link can still be maintained. It is worth noticing that the \ac{bse} encountered in \acp{elaa} and \acp{ris} can be eliminated in \ac{dmimo} {because of the digital implementation of the system (i.e., one radio frequency chain for each antenna {widely distributed in space}).

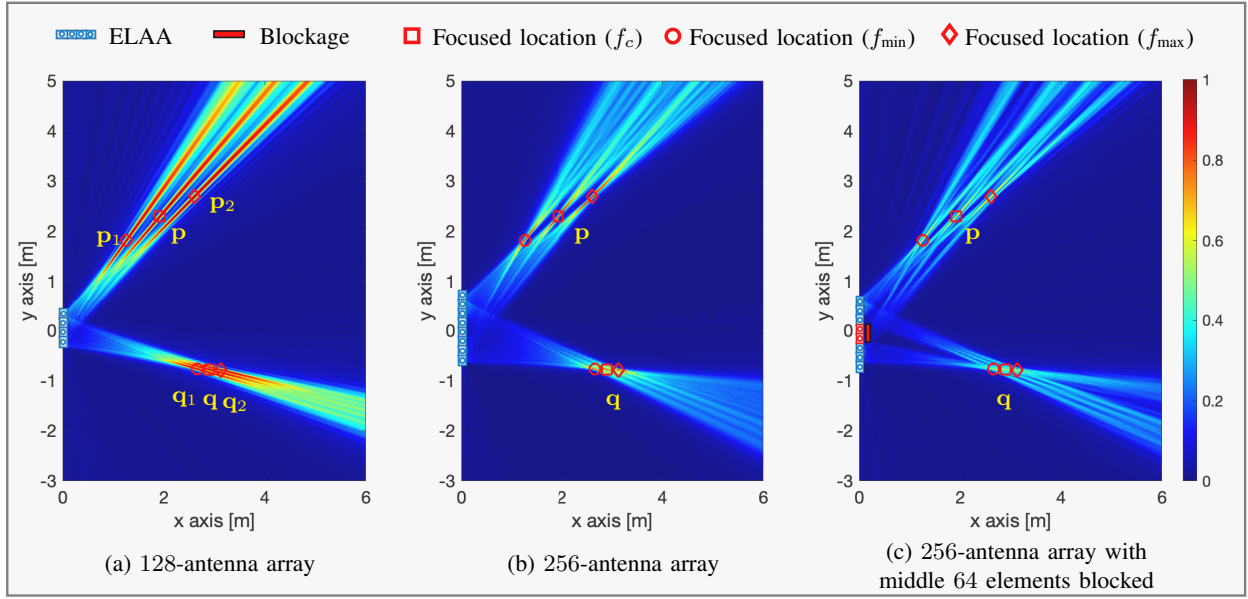
\begin{figure*}[t]
\centering
\begin{tikzpicture}
\node (image) [anchor=south west]{\includegraphics[width=.91\linewidth]{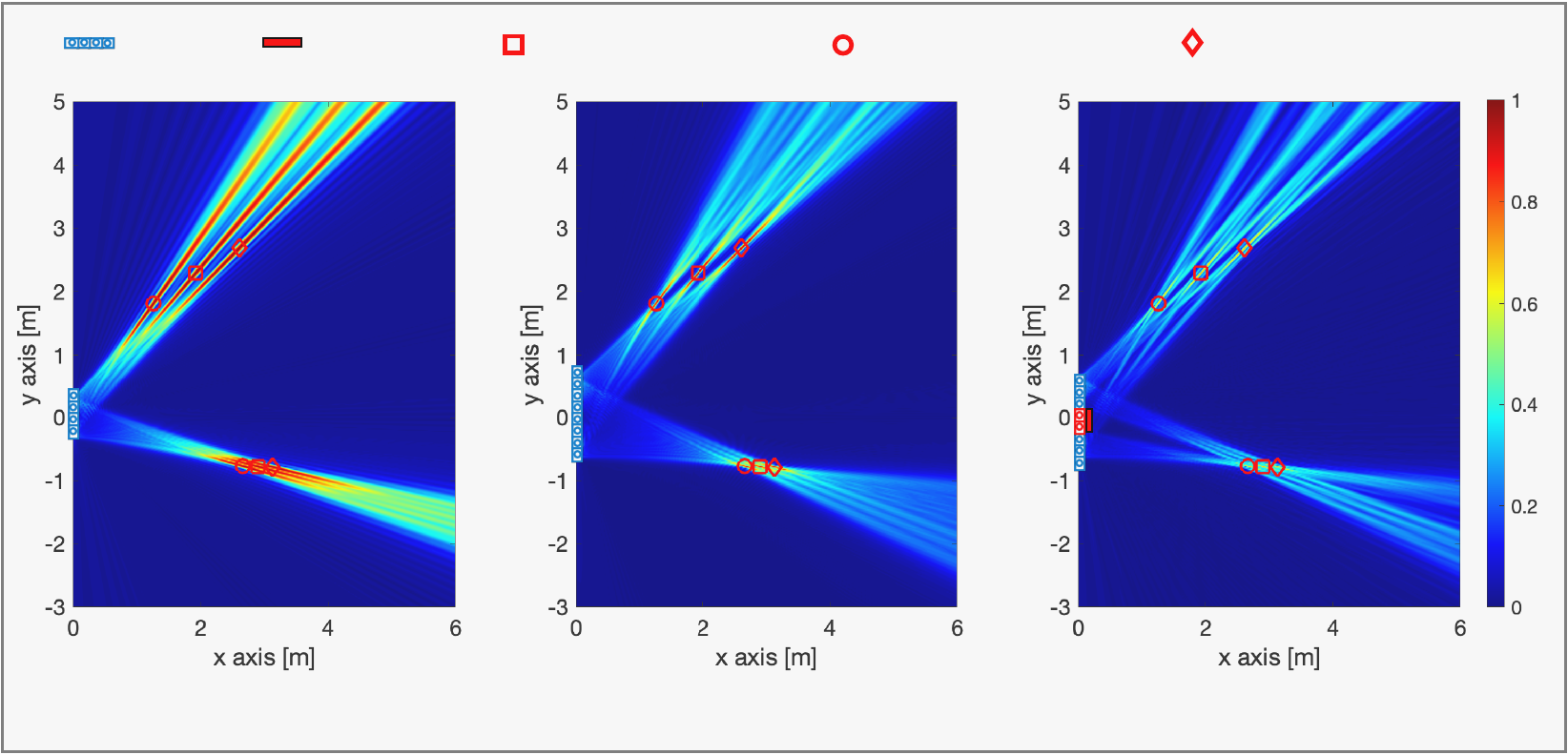}};
    \gettikzxy{(image.north east)}{\ix}{\iy};
\node at (0.09*\ix,0.625*\iy)[rotate=0,anchor=north]{\small{\yellow{$\pv_1$}}};
\node at (0.145*\ix,0.63*\iy)[rotate=0,anchor=north]{\small{\yellow{$\pv$}}};
\node at (0.18*\ix,0.68*\iy)[rotate=0,anchor=north]{\small{\yellow{$\pv_2$}}};
\node at (0.15*\ix,0.365*\iy)[rotate=0,anchor=north]{\small{\yellow{$\qv_1$}}};
\node at (0.17*\ix,0.36*\iy)[rotate=0,anchor=north]{\small{\yellow{$\qv$}}};
\node at (0.19*\ix,0.355*\iy)[rotate=0,anchor=north]{\small{\yellow{$\qv_2$}}};
\node at (0.465*\ix,0.63*\iy)[rotate=0,anchor=north]{\small{\yellow{$\pv$}}};
\node at (0.49*\ix,0.36*\iy)[rotate=0,anchor=north]{\small{\yellow{$\qv$}}};
\node at (0.775*\ix,0.63*\iy)[rotate=0,anchor=north]{\small{\yellow{$\pv$}}};
\node at (0.80*\ix,0.36*\iy)[rotate=0,anchor=north]{\small{\yellow{$\qv$}}};

\node at (0.115*\ix,0.955*\iy)[rotate=0,anchor=north]{\small{ELAA}};
\node at (0.245*\ix,0.955*\iy)[rotate=0,anchor=north]{\small{Blockage}};
\node at (0.43*\ix,0.955*\iy)[rotate=0,anchor=north]{\small{Focused location ($f_c$)}};
\node at (0.64*\ix,0.955*\iy)[rotate=0,anchor=north]{\small{Focused location ($f_\text{min}$)}};
\node at (0.86*\ix,0.955*\iy)[rotate=0,anchor=north]{\small{Focused location ($f_\text{max}$)}};
\node at (0.17*\ix,0.10*\iy)[rotate=0,anchor=north]{\small{(a) $128$-antenna array}};
\node at (0.49*\ix,0.10*\iy)[rotate=0,anchor=north]{\small{(b) $256$-antenna array}};
\node at (0.81*\ix,0.12*\iy)[rotate=0,anchor=north]{\small{\shortstack{(c) $256$-antenna array with\\ middle $64$ elements blocked}}};
\end{tikzpicture}
\caption{Illustration of NF channel features {and normalized signal power measurements at different locations} ({$f_\text{c}=$} $\unit[30]{GHz}$ system with 4 GHz bandwidth). 
{In each scenario, a {half-wavelength spaced} \ac{elaa} is beamforming to two \acp{ue}, located at $\pv = [1.928, 2.298]^\top\unit[]{m}$ (angle-of-departure $\theta_\pv=50^\circ$) and $\qv = [2.898, -0.777]^\top\unit[]{m}$ ($\theta_\qv=-15^\circ$), both are $\unit[3]{m}$ away from the array center. The received signal power is normalized by the amplitude of the channel gain and the number of antennas (scaled from 0 to 1). Due to the \ac{bse}, the focused locations of the lowest subcarrier ($f_\text{min} = \unit[28]{GHz}$) and highest subcarriers ($f_\text{max} = \unit[32]{GHz}$) are actually $\pv_1=[1.263, 1.815]^\top\unit[]{m}, \qv_1=[2.661,-0.768]^\top$\unit[]{m} and $\pv_2=[2.610, 2.693]^\top\unit[]{m}, \qv_2=[3.131,-0.783]^\top\unit[]{m}$, respectively (the other subcarriers are not visualized).}
}
\vspace{-5mm}
\label{fig-2-nf_features}
\end{figure*}

\begin{table*}[ht]
\centering
\scriptsize
    \caption{{{A Summary of Near-Field-Enabling Scenarios}}}
    \centering
    \renewcommand{\arraystretch}{1.2}
    \begin{tabular}{c !\vthickline c | m{2.0cm} | m{5.2cm} | m{3.4cm} | m{4.2cm}}
    \hthickline
     \multicolumn{2}{c|}{\!\!\textbf{Tasks}} 
    & \makecell{\textbf{Scenarios}}
    & \makecell{\textbf{FF}} 
    & \makecell{\textbf{NF}}
    & \makecell{\textbf{NF Benefits}}
    \\
    \hthickline
        \multirow{7}{*}{\rotatebox{90}{Localization}} 
        & \cellcolor{blue!5} 
        L1 
        & \cellcolor{blue!5}
        Location estimation of a single-antenna UE (without RIS) 
        & \cellcolor{blue!5}
        {a) Time-difference-of-arrival (multiple single-antenna BSs); \newline
        b) Angle-based (multiple multi-antenna BSs, or single multi-antenna BS with absolute range measurement via large bandwidth and round-trip time protocols).}
        & \cellcolor{blue!5}
        a) Phase coherence exploitation in D-MIMO; \newline
        b) SWM exploitation of a single BS, without the wideband requirement.
        & \cellcolor{blue!5}
        - Reduced bandwidth (with SWM exploitation using narrowband signals); \newline
        - Reduced number of BSs or {elimination of} absolute range requirement for angle-based methods (position-dependent array response).
        \\
        \hhline{~|*{5}{-}} 
        & \cellcolor{blue!5}
        L2
        & \cellcolor{blue!5}{Location estimation of a single antenna UE (with RIS)} 
        & \cellcolor{blue!5}{
        a) Single-RIS-enabled localization (LOS and wideband required); \newline
        b) Multi-RIS-enabled sidelink localization in out-of-coverage scenarios.} & \cellcolor{blue!5}{
        a) Work with LOS blockage (and possibly narrowband signals);\hspace{2cm}
        b) Work with single-RIS and narrowband signals.}
        & \cellcolor{blue!5} 
        - Extended detection area (LOS blockage); \newline
        - Reduced bandwidth (with SWM); \newline
        - Reduced number of RIS anchors.
        \\ 
        \hhline{~|*{5}{-}} 
        & \cellcolor{blue!5} 
        L3 
        & \cellcolor{blue!5}
        Location and orientation estimation of a multi-antenna UE
        & \cellcolor{blue!5}
        a) At least 2 multi-antenna anchors (BSs or RISs) for 6D localization; \newline
        b) Single-anchor with NLOS paths.
        & \cellcolor{blue!5}{
        a) Work with single-anchor and narrowband signals; \newline
        b) NLOS paths are not required.}
        & \cellcolor{blue!5} 
        - Reduced number of anchors; \newline
        - Reduced bandwidth.
        \\
        \hthickline
        \multirow{4}{*}{\rotatebox{90}{Sensing\ }} 
        & \cellcolor{yellow!10}
        S1
        & \cellcolor{yellow!10}
        Bistatic/multistatic sensing of a passive target
        & \cellcolor{yellow!10}
        a) ELAA/D-MIMO aided sensing; \newline
        b) RIS-aided sensing (possibly with double- or multi-bounce channels). \newline
        (both require wideband for location estimation)
        & \cellcolor{yellow!10}
        a), b) Targets located at the same AOA could be resolved with beamfocusing, and work with narrowband signals.
        & \cellcolor{yellow!10}
        - Extended detection area; \newline
        - Reduced bandwidth (with SWM).
        \\ \hhline{~|*{5}{-}} 
        & \cellcolor{yellow!10}
        S2
        & \cellcolor{yellow!10}
        Sensing of an RIS target
        & \cellcolor{yellow!10}
        Multistatic sensing of a target equipped with an RIS. (at least 3 anchors are needed, e.g., 1 transmitter, 2 receivers)
        & \cellcolor{yellow!10}
        Work with a bistatic sensing setup (e.g., 1 transmitter, 1 receiver).
        & \cellcolor{yellow!10}
        - Reduced number of anchors required.
        \\ 
    \hthickline
    \multirow{4}{*}{\rotatebox{90}{Joint L\&S \ }} 
        & \cellcolor{red!5}
        J1
        & \cellcolor{red!5}
        Joint L\&S of UE and passive targets (without RIS)
        & \cellcolor{red!5}
        a) Multi-antenna BS and multi-antenna UE; \newline
        b) Multi-antenna BS and single-antenna UE (SIMO) with Doppler estimation (mobile UE).
        & \cellcolor{red!5}
        a) Work with SIMO/MISO systems;
        \newline
        b) Work with stationary scenarios.
        & \cellcolor{red!5}
        - Reduced hardware cost (multi-antenna not required); \newline
        - Extended application scenarios (no mobility required).
        \\ \hhline{~|*{5}{-}}
        & \cellcolor{red!5}
        J2
        & \cellcolor{red!5}
        Joint RIS calibration and UE positioning
        & \cellcolor{red!5}
        6D RIS calibration and 3D UE positioning with a multi-antenna BS and multiple single-antenna UEs. (or a single UE at multiple locations).
        & \cellcolor{red!5}
        Work with a single UE at a fixed location.
        & \cellcolor{red!5}
        - Extended application scenarios (multiple UEs are not required).
        \\
    \hthickline
    \end{tabular}
    \vspace{0.1 cm}
    \scriptsize{\raggedright \\ *Notes: The abbreviations `L', `S', and `J' denote `localization', `sensing', and `joint localization and sensing', respectively. {The \ac{las} systems discussed in this table are in uplink; however, tasks can also be done downlink with different protocols. In addition to enabling scenarios, NF can also boost the \ac{las} performance, which is not discussed in this table.}
    \par}
    \renewcommand{\arraystretch}{1}
    \label{tab:summary_of_enabling_scenarios}
\end{table*}

\section{Opportunities for Localization and Sensing}
\label{sec_opportunities_for_las}

\ac{nf} operation introduces several distinct features that are not present in \ac{ff} operation {to boost communications}, as discussed in the previous section. These features also bring opportunities for \ac{las}, which will be detailed in this section {with NF-enabling scenarios summarized in Table~\ref{tab:summary_of_enabling_scenarios}.}

\subsection{{NF L\&S vs. NF Communication}}
Wireless communication and L\&S systems can be implemented in the same hardware and be configured to use the same waveforms and signal processing blocks; however, they target different goals.
In particular,
communication systems are designed to transfer information through
a radio channel acting as a \emph{medium}, resulting in treating the \emph{aggregate}
channel as a nuisance in order to optimize data transmission. In contrast, L\&S systems are designed with the sole purpose of extracting information about the UE (or passive target) state from that radio channel. As a result, \ac{las} benefit from the NF features in a manner distinct from that of communication systems.

In L\&S systems, the coupling in range and angle coming from the \ac{swm} allows for direct localization of both users and scatterers with high accuracy. In this case, the range-based localization (requiring wideband signals) and angle-based localization (requiring multiple multi-antenna BSs) in the FF are no longer required, thus enabling frugal \ac{las} with narrowband signals~\cite{dardari2022nlos}.
Similarly, beamfocusing in the NF can be utilized to spatially resolve closely spaced scatterers {for high-resolution radar sensing, which cannot be done with beamforming in the FF.} 
While SNS does not require extra effort in communications as the diversity gain in the end-to-end channel is elegantly exploited, \ac{las} must accurately model this feature, especially for geometric parameter extraction of extended targets. This effect can also be leveraged to detect channel {blockages,} variations and potentially enable passive target sensing. 
Regarding the \ac{bse}, L\&S can further exploit location information {contained in the squinted beams formed by different subcarriers, thus improving the L\&S accuracy and resolution with beamfocusing with limited scanning overhead~\cite{cui2022rainbow}}.

\subsection{NF L\&S: ELAA vs. RIS vs. D-MIMO}

ELAAs and \ac{dmimo} are usually treated as separate architectures; however, the particular distinction {in terms of NF \ac{las}} is not black and white. For example, an ELAA can be viewed as a special case of a D-MIMO system when the array is large enough, as in~\cite{nearfieldSense_TWC_2022}. {Similarly, a phase-coherent D-MIMO system can be treated as an ELAA.} The distinction is that ELAAs are typically composed of dense antenna arrays in a uniform structure that are directly connected to a central processing unit, while D-MIMO systems consist of sparse and widely separated antennas that are connected through a shared fronthaul to potentially several processing units.

The NF features experienced in RIS-aided systems are similar to those in ELAAs. {In the LOS-blocked scenario, the BS receives signals from the UE through the RIS channel with time-varying RIS profiles/codebooks, which is analogous to the combiner change of an analog ELAA. The NF features introduced by the RIS are beneficial to \ac{las}. If the LOS path is available, extra geometrical information {becomes accessible}, but new challenges arise in beam design and channel parameter extraction. Compared to active devices such as ELAAs and D-MIMO, the nearly passive and low-cost properties of the RISs yield a more practical deployment of systems operating in the NF.}
In the following, three localization scenarios, two sensing scenarios, and two joint scenarios are discussed as examples of NF \ac{las} systems.

\subsection{{NF Localization}}

\subsubsection{L1-Location Estimation without RIS}\label{sec_l1}
{Time-difference-of-arrival-based localization is widely used to locate a single-antenna UE (L1-(a))}. In such a system, multiple time-synchronized \ac{bs} cooperate to localize a \ac{ue} through range-based measurements. With the introduction of phase-synchronization and phase-coherent processing, which can
especially be provided at sub-6 GHz frequencies (e.g., via an all-digital sigma-delta-over-fiber architecture), 
the entire D-MIMO system can be turned into a very large sparse array. {By exploiting the SWM property of the signal}, narrowband localization of \acp{ue} using carrier phase only becomes possible~{\cite{fascista2023uplink}}. Similarly, angle-based localization (L1-(b)) can also benefit from the NF features by exploiting SWM.
Centimeter level localization accuracy has been achieved with \ac{nf} D-MIMO- or ELAA-based localization, using local or centralized implementations methods~\cite{de2022expert}.

\begin{figure}[t]
\centering 
\centerline{
%
%
\definecolor{mycolor1}{rgb}{0.00000,0.45000,0.74000}%
\definecolor{mycolor2}{rgb}{0.85000,0.33000,0.10000}%
\definecolor{mycolor3}{rgb}{0.93000,0.69000,0.13000}%
\definecolor{mycolor4}{rgb}{0.49000,0.18000,0.56000}%

\begin{tikzpicture}

\begin{axis}[%
width=72mm,
height=40mm,
at={(0mm,0mm)},
scale only axis,
xmin=0.1,
xmax=1000,
xmode=log,
ymin=0.001,
ymax=1,
ymode=log,
yticklabel style = {font=\small,xshift=0.5ex},
xticklabel style = {font=\small,yshift=0ex},
axis background/.style={fill=white},
axis background/.style={fill=white},
xmajorgrids,
ymajorgrids,
legend style={font=\scriptsize, at={(0.4, 0.12)}, anchor=south west, legend cell align=left, align=left, draw=white!15!black, style={row sep=-0.1cm}}
]

\addplot [color=mycolor1, line width=1.5pt]
  table[row sep=crcr]{%
0.1	0.278123956172868\\
0.193069772888325	0.278041896166843\\
0.372759372031494	0.277853531020423\\
0.719685673001151	0.277634898594699\\
1.38949549437314	0.277404701273964\\
2.68269579527973	0.276693409217517\\
5.17947467923121	0.27501085595482\\
10	0.270814284848842\\
19.3069772888325	0.260713225489979\\
37.2759372031494	0.239226464161434\\
71.9685673001153	0.202141418974397\\
138.949549437314	0.149306232001105\\
268.269579527973	0.0854665547548432\\
517.947467923122	0.0436310770202234\\
1000	0.0228015085858023\\
};
\addlegendentry{\footnotesize{Noncoherent ($M = 4$)}}

\addplot [color=mycolor2, dotted, line width=1.5pt]
  table[row sep=crcr]{%
0.1	0.135723789003669\\
0.193069772888325	0.135714248946037\\
0.372759372031494	0.135692323998916\\
0.719685673001151	0.135666830276377\\
1.38949549437314	0.135639934737297\\
2.68269579527973	0.135556482716799\\
5.17947467923121	0.135356971347581\\
10	0.134846133136485\\
19.3069772888325	0.133534368067797\\
37.2759372031494	0.130300884953867\\
71.9685673001153	0.122858592034906\\
138.949549437314	0.105917802903703\\
268.269579527973	0.0699769091924819\\
517.947467923122	0.0391298387956423\\
1000	0.0218706195328406\\
};
\addlegendentry{\footnotesize{Noncoherent ($M = 8$)}}

\addplot [color=mycolor3, dashdotted, line width=1.5pt]
  table[row sep=crcr]{%
0.1	0.0674630692765006\\
0.193069772888325	0.06746189756523\\
0.372759372031494	0.0674592039824899\\
0.719685673001151	0.0674560706309059\\
1.38949549437314	0.0674527634377227\\
2.68269579527973	0.0674424916801414\\
5.17947467923121	0.0674178724077019\\
10	0.0673544335231638\\
19.3069772888325	0.0671888382783038\\
37.2759372031494	0.0667634554663021\\
71.9685673001153	0.065682529160103\\
138.949549437314	0.0625635634921251\\
268.269579527973	0.0509999697342851\\
517.947467923122	0.0325300482260354\\
1000	0.0196231747938589\\
};
\addlegendentry{\footnotesize{Noncoherent ($M = 16$)}}

\addplot [color=mycolor4, dashed, line width=1.5pt]
  table[row sep=crcr]{%
0.1	0.00189085384527453\\
0.193069772888325	0.0018908289480405\\
0.372759372031494	0.0018907808755791\\
0.719685673001151	0.00189068809440641\\
1.38949549437314	0.00189050902133383\\
2.68269579527973	0.00189016333372772\\
5.17947467923121	0.00188949621429584\\
10	0.00188820934427691\\
19.3069772888325	0.0018857292057521\\
37.2759372031494	0.00188095784763132\\
71.9685673001153	0.00187181019595134\\
138.949549437314	0.00185438165479281\\
268.269579527973	0.0018214865910792\\
517.947467923122	0.00176055884895266\\
1000	0.00165240579578259\\
};
\addlegendentry{\footnotesize{Coherent ($M=4,8,16$)}}

\end{axis}

\draw [stealth-stealth] (0.3, 0.4) -- (0.3, 2.4);

\node[rotate=0] at (1.6cm, 1.4cm) {\small{\shortstack[l]{Accuracy gain\\ via phase coherent\\processing}}};

\node[rotate=0,fill=white] 
at (3.75cm,-.75cm){\small Bandwidth [MHz]};
\node[rotate=90] 
at (-10mm,20mm){\small Position Error Bound [m]};
\end{tikzpicture}%
}
\caption{Position error bound vs. bandwidth for uplink localization of a single-antenna UE in a distributed \acp{bs} network at $\unit[3.5]{GHz}$, consisting of $4$ \acp{bs} equipped with $M$-element uniform linear array. The \acp{bs} are deployed on the corners of an area of size $\unit[10 \times 10]{m^2}$ at the height of $\unit[3]{m}$, while the UE is located at $[2, 9, 1]^\top \unit[]{m}$. The position error bound values refer to 2D localization accuracy under the assumption of known UE height {and are computed using \cite[Eq.~(41)]{fascista2023uplink}}. In the coherent scenario, \acp{bs} are phase synchronized, creating a spatially distributed ELAA working in the NF. Simulation details and other parameters can be found in \cite[Sec.~V-A]{fascista2023uplink}.
}
\label{fig-6}
\end{figure}
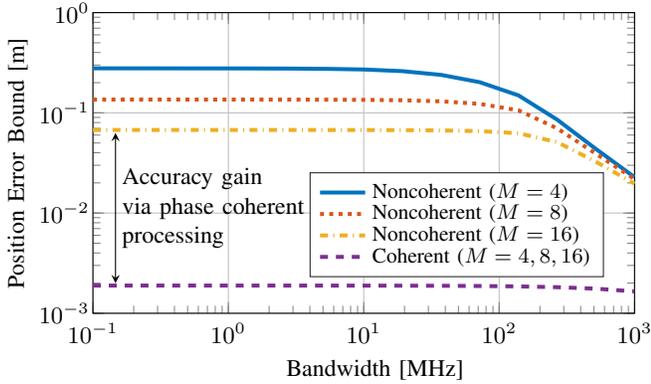

\emph{Case Study 1:}
Fig.~\ref{fig-6} shows the uplink localization performance of a single-antenna UE in a phase-coherent network {(L1-(a))} with and without phase synchronization among the distributed \acp{bs}~\cite{fascista2023uplink}. Phase-synchronized (-coherent) scenario corresponds to NF operation since the D-MIMO network seen as a whole can be treated as a large aperture array created by a combination of distributed phase-coherent \acp{bs}. On the other hand, the absence of phase synchronization (non-coherent {scenario}) prevents exploitation of the \ac{nf} features (e.g., SWM). We observe that phase-coherent processing (enabled in the coherent scenario) can provide significant gains in localization accuracy over the non-coherent scenario. In addition, accuracy in the coherent case stays almost constant with respect to bandwidth since the carrier phase conveys much richer delay/position information than {the one provided through bandwidth (i.e., sharper peaks around the true UE location in the likelihood function).}

\subsubsection{{L2-Location Estimation with RIS}}
\Acp{ris}, working as reference anchors, can enable localization (even for the simplest \ac{siso} scenario) with wideband signals due to the extra angular information {(L2-(a))}. More specifically, the position of a \ac{ue} can be obtained as the intersection of the line from \ac{ris} to UE (corresponding to {angle-of-departure} estimate) and a hyperboloid formed by two delay estimates from the LOS path and the \ac{ris} path. In the \ac{nf} scenario, this system can work with narrowband signals and under LOS blockage (which limits the localizability in the \ac{ff} cases) by exploiting the \ac{swm} feature of the signal. {This scenario is similar to L1-(b) by replacing an ELAA with a single antenna transmitter/receiver working together with an RIS. In addition, NF features can provide more L\&S opportunities in sidelink localization (L2-(b)) in out-of-coverage areas, where more details describing the scenarios can be found in~\cite{chen2023riss}.}

\emph{Case Study 2:} A numerical example of RIS-assisted NLOS localization is shown in Fig. \ref{fig_ris_loc}, where a single-antenna \ac{ue} is navigating in an indoor environment. A long linear \ac{ris} is deployed, and the \ac{ue} can localize itself by analyzing the signal reflected by the \ac{ris} even under LOS blockage scenarios~\cite{dardari2022nlos}. Despite the presence of many obstacles that partially obstruct the \ac{ris}, the localization error along the \ac{ue} trajectory remains limited to $\unit[20-30]{cm}$, as seen from the inset plot. Note that wideband signal is not a requirement but can facilitate localization tasks with extra time measurements.

\begin{figure}[t]
\centering
\begin{tikzpicture}
\node (image) [anchor=south west]{\includegraphics[width=1\linewidth]{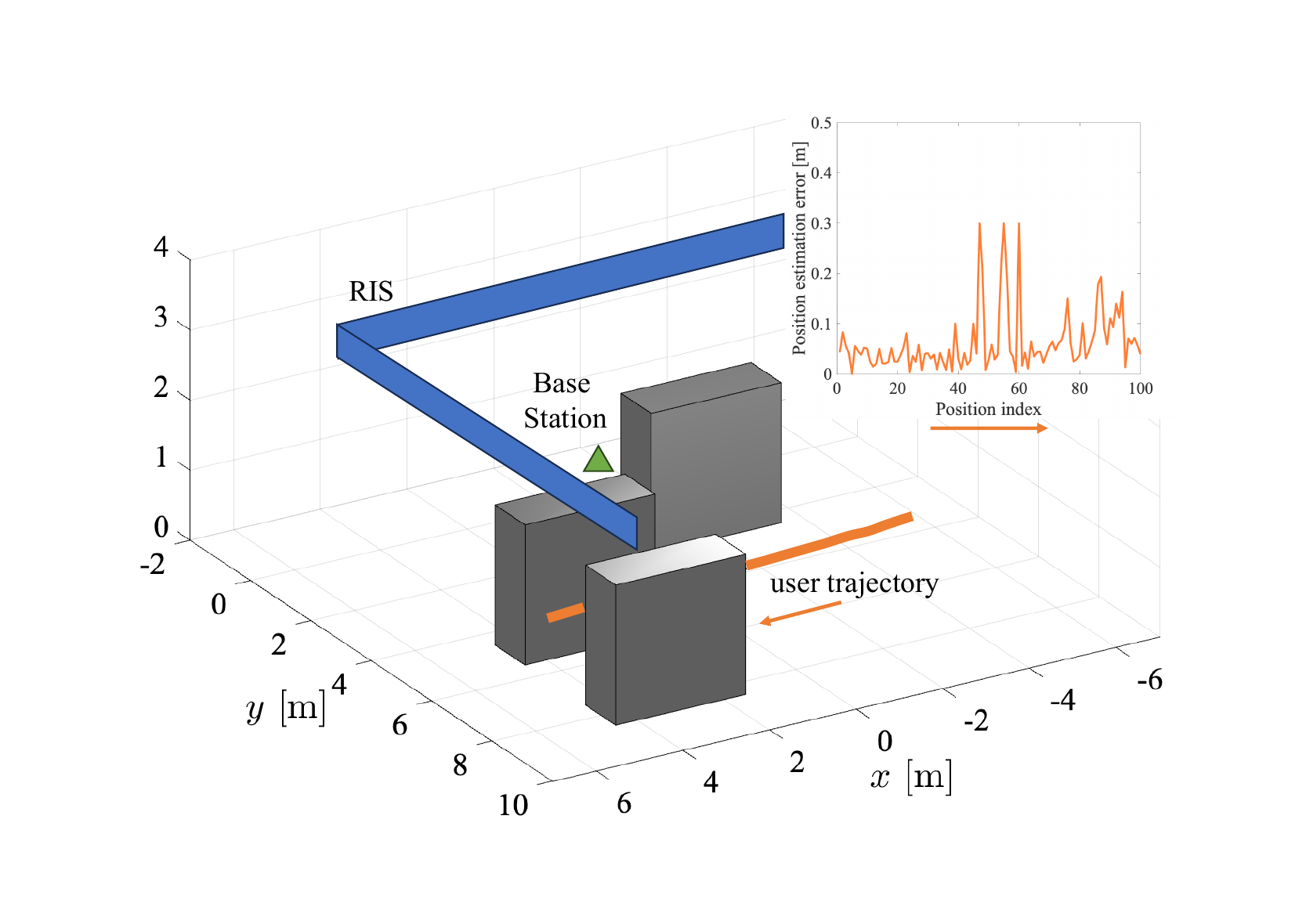}};
\end{tikzpicture}
\caption{RIS-assisted localization in a {NF} NLOS scenario {{(L2-(a))}} ($\unit[28]{GHz}$ system with $\unit[250]{MHz}$ bandwidth). A single-antenna UE {moves} along the trajectory {(from $[-4, 6,1]^\top\unit[]{m}$ to $[4, 6,1]^\top\unit[]{m}$) (orange line)}, and a BS in $[0, 0,1]^\top\unit[]{m}$ (green triangle) 
sends positioning reference signals, which are reflected by {a $20\unit[]~{m}$ long linear RIS deployed from $[-5, 0,3]^\top\unit[]{m}$ to $[5, 0,3]^\top\unit[]{m}$ and from $[5, 0,3]^\top\unit[]{m}$ to $[5, 10,3]^\top\unit[]{m}$ (blue areas). The error evolution along the trajectory is reported; larger errors} occur around index 50 due to the blockage of a large amount of RIS elements. {Time-varying RIS reflection coefficients are exploited to discriminate between paths coming from different portions of the RIS, and compute a set of BS-RIS-UE delay measurements. Based on these delay measurements, the UE position is computed, even in the case of obstructed BS-UE path \cite{dardari2022nlos}.}}
\label{fig_ris_loc}
\end{figure}

\subsubsection{{L3-Location and Orientation Estimation of a Multi-antenna UE}}
When considering 6D localization (i.e., 3D position and 3D orientation) of \acp{ue} in the \ac{ff}, at least 2 anchors (BSs or RISs with known state information) are needed for 3D orientation estimation of a UE {(L3-(a))}. {Specifically, after position estimation via the intersection of two direction vectors originating from the anchors, the rotation matrix is calculated based on the two estimated local angle pairs with respect to the two anchors at the UE.}
In the presence of \ac{nlos} paths, multipath exploitation can enable 6D \ac{ff} localization with a single {anchor (L3-(b))}.
By leveraging the \ac{swm} feature in the \ac{nf} (and possibly \ac{sns}/\ac{bse}, which also {convey} the position and orientation information of the UE), it becomes possible to achieve single-{anchor} 6D localization without requirements on the propagation environment (i.e., the existence of NLOS paths).

\subsection{{NF Sensing}}
\subsubsection{{S1-Bistatic/Multistatic Sensing of a Passive Target}}
The key advantage of ELAA and D-MIMO from a monostatic/bistatic sensing perspective is the dense and widely distributed deployment of antennas, which provides good coverage, high accuracy, and improved resolution {(S1-(a))}. A particular mode of operation is coherent processing, which has the potential of achieving remarkable spatial resolution even at narrow bandwidths \cite{nearfieldSense_TWC_2022}. 
In terms of radar integration, the sensing can either be integrated in the downlink or in the uplink, with carefully selected antennas as transmitters or receivers.
In each case, the key objective is to integrate the radar functionality into the communication system at minimum communication overhead and loss in radar performance. 
Regarding \ac{ris}-aided sensing {(S1-(b))}, most of the existing literature focuses on \ac{ff} scenarios to improve sensing performance \cite{foundations_RIS_radar_TSP_2022}. Compared with the traditional sensing system without \ac{ris}, the quality of the reflected signal from the target can be improved, and the sensing service can be available with LOS blockage. In addition, an \ac{ris} located far away from the radar virtually extends the aperture of the system (or can be interpreted as an extra radar).
When operating in the \ac{nf}, similar to {(S1-(a))}, \ac{ris} can {help} resolve targets residing in the same angle, but at different distances, through \ac{swm}.

\emph{Case Study 3:} An example of NF sensing experimental setup and measurement results are shown in Fig.~\ref{fig_dmimo_setup}, using the same setup but different signal frequency as~\cite{nearfieldSense_TWC_2022}. The heatmaps illustrate the matched filter response of a moving cylinder using an experimental massive MIMO testbed configured for bistatic radar sensing, emulating an ELAA or D-MIMO system operating in the NF. In addition to the high resolution, the experimental result also highlights some challenging non-idealities (e.g., hardware impairments) that are common in the NF of large antenna arrays, which have the impact of introducing mainlobe and sidelobe distortions in the radar image.

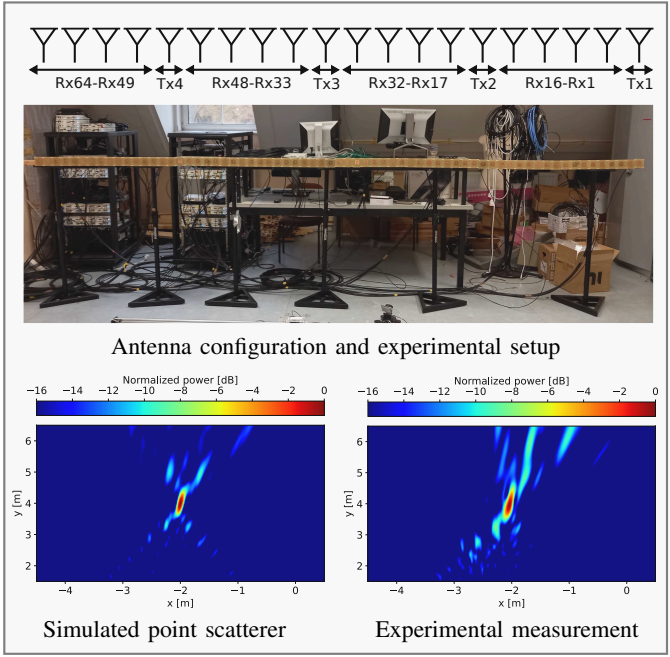
\begin{figure}[t]
\centering
\begin{tikzpicture}
\node (image) [anchor=south west]{\includegraphics[width=1\linewidth]{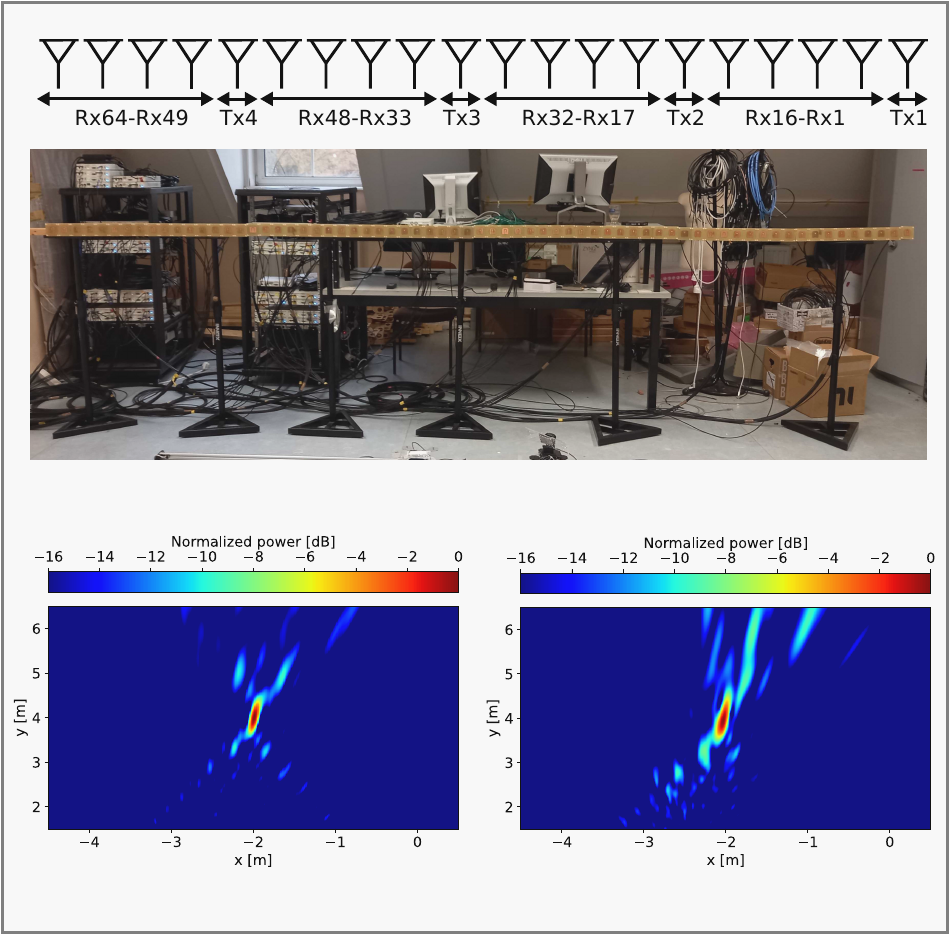}};
    \gettikzxy{(image.north east)}{\ix}{\iy};
\node at (0.5*\ix,0.50*\iy)[rotate=0,anchor=north]{\small{Antenna configuration and experimental setup}};
\node at (0.25*\ix,0.08*\iy)[rotate=0,anchor=north]{\small{Simulated point scatterer}};
\node at (0.75*\ix,0.08*\iy)[rotate=0,anchor=north]{\small{Experimental measurement}};
\end{tikzpicture}
\caption{NF ELAA/D-MIMO sensing experimental setup and measurement results. {The testbed (with a length of $68\times \unit[7]{cm} = \unit[4.76]{m}$) operates at $\unit[2.61]{GHz}$ using the OFDM waveform with $100$ subcarriers and $\unit[18]{MHz}$ bandwidth (which corresponds to $(\unit[3\times 10^8]{m/s})/(\unit[18\times 10^6]{Hz}) = \unit[16.67]{m}$ in bistatic range resolution). The system was calibrated using a reference measurement with the cylinder ($\unit[25]{cm}$ in diameter, and the size is not taken into account in the reconstruction) at a known location for phase and amplitude alignment. Localization is performed with the cylinder moving at around $\unit[1]{m/s}$ (larger than the bistatic Doppler resolution of $\unit[0.1149]{m}/\unit[0.2]{s}=\unit[0.5745]{m/s}$). The data processing is carried out on the channel state information using range-Doppler processing and static removal, followed by an image
formation using the back projection.} As can be seen,
a high spatial resolution is obtained in the NF, despite a narrow bandwidth. {More details can be found in~\cite{nearfieldSense_TWC_2022}}.
}
\vspace{-5mm}
\label{fig_dmimo_setup}
\end{figure}

\subsubsection{{S2-Sensing of an RIS Target}}
An RIS attached to a target can provide a much stronger reflected signal and orientation estimation of itself. In the \ac{ff}, at least 3 anchors (e.g., 1 transmitter and 2 receivers) are needed to localize the target equipped with an \ac{ris} {(given each transmitter-RIS-receiver path provides a delay estimation and 2D spatial frequency estimation).} In the \ac{nf}, however, a single-antenna transmitter and a single-antenna receiver are sufficient for localization in a bistatic sensing setup. Such sensing scenarios can also be interpreted as calibrating \ac{ris} anchors with known transmitter and receiver states.

\subsection{{NF Joint Localization and Sensing}}

\subsubsection{{J1-Joint L\&S of UE and Passive Targets}}
Sensing can also be performed jointly with localization tasks, especially when the surrounding environment is dynamic. For example, the positions of incident points (that create extra paths between the transmitter and receiver) can also be estimated as a by-product in simultaneous localization and mapping scenarios. In the \ac{ff}, this can be accomplished with a multiple-antenna UE ({J1-(a)}), or a single-antenna UE under mobility by exploiting Doppler information~\cite{han2015performance} ({J1-(b)}). {In the \ac{nf}, the \ac{sns} (passive target sensing) and \ac{swm} features can be exploited to perform \ac{las} (of both reflecting and blocking passive targets) using a single-antenna UE, without the assistance of Doppler estimation.}  
The opportunities provided {by} RIS-aided joint localization and sensing are similar, which are not detailed here.

\subsubsection{{J2-Joint RIS Calibration and UE Positioning}}
For the RIS-aided \ac{las} system, calibration of the RIS is a crucial and practical issue that could be solved using the same method as {S2}. However, the state information of the transmitter(s) and receiver(s) are required. A more practical process is jointly performing UE positioning and RIS calibration tasks ({J2}), which requires a multi-antenna BS. For a single-antenna UE at a fixed location, two delay estimates, four angle estimates at the BS, and two spatial frequency estimates at the RIS can be obtained, which are insufficient for the calibration task. {In order to obtain the full 6D state of the \ac{ris} and 3D position of the UE in the \ac{ff} scenarios, the measurements from multiple single-antenna UEs or a single-antenna UE at different locations are needed (as one location provides). In contrast, this task can be tackled in the \ac{nf} with {one} single-{antenna} \ac{ue}, largely reducing the operation cost.

\subsection{{Summary of NF Benefits for L\&S}}
{Based on the described scenarios and systems with different technologies, the benefits of \ac{las} in the NF are summarized as follows:
\begin{itemize}
    \item \textbf{A more sustainable \ac{las} system}: remove the requirements of multi-antenna arrays (J1); reduce the number of anchors (L1, L2, L3, S2); support narrowband \ac{las} instead of the wideband requirement in the FF (L1, L2, L3, S1);
    spatial focusing to allow accurate sweeping-based \ac{las} due to the native high spatial resolution; the reduced radio resources in high accuracy L\&S allow a better coexistence with communication functionalities.
    \item \textbf{Extended \ac{las} application scenarios}: joint localization and sensing without mobility requirement (J1); joint RIS calibration and UE positioning without multi-UE requirement (J2){; possibility to detect occluded targets (S1); provide localization services in the areas with LOS blockage (L2).} 
    \item \textbf{Increased accuracy and trustworthiness:} improve \ac{las} accuracy compared to \ac{ff} counterparts,  when properly harnessing the \ac{nf} effects  (L1--J2); novel ways of utilizing \ac{nf} Doppler provide improved tracking performance; spatial focusing leads to natural interference mitigation as well as improved privacy.
\end{itemize}}

\section{Open Research Challenges}
In addition to the opportunities brought by the \ac{nf} features, it is essential not to overlook the challenges faced by \ac{las},
which are discussed in this section.

\subsection{Efficient NF Algorithms}
\label{sec:NFalg}
Most traditional localization algorithms have been developed for \ac{ff} configurations and follow a pragmatic two-step approach (i.e., extracting intermediate geometric parameters such as delay and \ac{aoa} followed by position estimation). In the \ac{nf}, due to the \ac{swm}, the maximum likelihood estimator can be adopted for direct positioning.
Alternative algorithms, such as compressive sensing, atomic norm minimization, cumulant-based algorithms, and two-step-style approximation are proposed. Regarding the sensing scenario, SNS requires extra effort in modeling for extended target localization and passive target sensing. In general, NF algorithms suffer from high complexity when adopting accurate channel models. However, using a simplified and approximated model (e.g., FF model) is sensitive to model mismatches (e.g., hardware impairments, partial blockage, diffraction of signals around objects) and degrades system performance.  The challenge lies in developing efficient NF algorithms, especially in scenarios where high accuracy and low latency are required.

\subsection{System Design and Optimization}
The design and optimization for \ac{las} systems are crucial, which can be classified into offline  (e.g., BS layout optimization) and online (e.g., frequency and power allocation) based on the key performance indicator requirements of the applications. Compared to the systems working in the \ac{ff}, \ac{swm} and \ac{sns} need to be considered in beamforming design, RIS profile optimization, and power allocation. Machine learning algorithms, such as reinforcement learning, can be good candidates to solve such high-complexity non-convex problems.
{Furthermore, concerns regarding electromagnetic field exposure may arise, requiring sensing technologies to detect and localize surrounding users to meet the energy constraints.} 
When considering the design of NF radar systems, processing schemes should be carefully crafted, taking into account the tradeoff between resolution and detection latency. In this regard, the characterization of extended targets, such as pedestrians or cars, remains an important research area.

\subsection{Misfocusing under Mobility}

The localization process utilizes ELAAs and RISs to focus the beam on a small spot centered around the UE. However, when the UE is in motion, there is a high likelihood that it will move outside this spot in the subsequent time slot. Consequently, the localization process and BS/RIS beamforming configuration must be restarted from scratch (or from surrounding areas of the position in the previous frames), causing link re-establish latency.
For sensing tasks, such misfocusing will cause the loss of target tracking.
The current state-of-the-art approach employing Bayesian filters is challenging to {apply} in the NF. First of all, the increased volume of measurements significantly increases system complexity.
Additionally, the observation model (e.g., from the antenna phase profile) also introduces strong nonlinearity, which can potentially cause the Bayesian filter to lose track of the user. 
{Considering the relatively small focused area in the NF, these challenges need to be overcome to ensure accurate UE and passive target tracking in a dynamic environment}.

\subsection{Hardware and System Level Challenges}

{There are plentiful hardware and system challenges in implementing the ELAAs, RISs, and D-MIMO into future systems, which could be the main bottleneck in getting the NF L\&S to work. While the main focus could be serving communication function only, for the multi-functional systems (e.g., \ac{las}), further efforts on hardware design and calibration algorithms are needed.}
In contrast to communication systems that only consider the end-to-end channel, dedicated calibration (e.g., antenna displacement and BS orientations) is a prerequisite for \ac{las} systems. Time synchronization and phase coherence across the antennas are important requirements, especially for \ac{dmimo} and \ac{elaa} scenarios. The time/phase errors are also affected by the calibration quality of the antenna positions at the system level for D-MIMO systems and the array level for ELAA and RIS-aided systems (e.g., element failures). In each case, techniques should be automatic, scalable, and cost-effective in order to ensure simple deployments with long life cycles. Another issue that may limit the sensing performance in D-MIMO systems is the fronthaul capacity, {potentially requiring data censoring and, consequently, compromising accuracy. As a remedy}, sequential signal processing strategies can be investigated for L\&S.

\subsection{Wideband NF} 
\ac{nf} \ac{las} rely on the variation of phase information across the \ac{elaa} and/or \ac{ris}. When wideband signals are considered {(e.g., in sub-THz and THz systems with large bandwidth)}, such variation is conditioned on the specific frequency component (e.g., subcarrier in the case of \ac{ofdm} implementation) due to the \ac{bse}. Consequently, different focal points can be experienced for different spectral components with a fixed phase configuration at the \ac{elaa}. However, the variation of the phase provides additional information concerning the source position. In this case, the position information comes from both the \ac{nf} effect and from the use of the wide bandwidth signal. Practical algorithms are foreseen to extract the positioning information in such a context with affordable complexity.

\subsection{Extended and Non-Isotropic Targets} 
Spatially extended and non-isotropic targets pose both modeling and signal processing challenges in NF \ac{las}. Isotropic reflectivity (often associated with point targets) means that the complex channel gain of a target (including the effects of radar cross-section, scattering-induced phase shift, and path attenuation) is the same for all transmit-receive antenna pairs \cite{nearfieldSense_TWC_2022}. However, non-isotropic scattering \cite{nearfieldSense_TWC_2022} happens in widely distributed and densely deployed \ac{dmimo} systems, as well as in scenarios where the target is sufficiently close to the antennas to make it spatially extended. {This is not because of the change of the target itself, but rather the NF condition makes the point-target model invalid for L\&S}. The extended and non-isotropic properties of targets {significantly impede} sensing applications {that employ} high-resolution phase-coherent processing algorithms, and {impair} localization applications {by introducing \ac{nlos} paths, complicating the localization process.}
Novel algorithms, such as tracking the complex scattering coefficients of targets across consecutive snapshots, leveraging adaptive signal processing techniques to extend the coherent aperture, and machine learning-based methods for target detection, can be developed.

\section{Conclusion}
{With the increased frequency and the adoption of large antennas in communication systems, NF scenarios are more likely to happen.} The \ac{nf} features affect communication, localization, and sensing differently, even though these services all share the same infrastructure and signal propagation medium. This is due to the fact that communication only considers the end-to-end channel, while \ac{las} have to extract geometric information from it. Regarding the opportunities brought {by} NF to \ac{las}, the SWM provides extra geometric information and enables a variety of application scenarios that are not possible in the FF condition. The SNS can contribute to passive target detection and localization based on channel variation, and BSE can accelerate the \ac{las} process by taking advantage of multiple focused areas at different subcarriers. However, to realize these benefits, accurate NF channel modeling and the design of corresponding signal processing and optimization algorithms are essential.


\bibliographystyle{IEEEtran}
\bibliography{ref}

\vspace{-0.5cm}
\begin{IEEEbiographynophoto}{Hui Chen}
(hui.chen@chalmers.se) is a postdoctoral researcher at Chalmers University of Technology, 41296 Gothenburg, Sweden. His research interests include mmWave/THz and RIS-aided localization. \end{IEEEbiographynophoto}

\vspace{-0.5cm}
\begin{IEEEbiographynophoto}{Musa Furkan Keskin}
(furkan@chalmers.se) is a research specialist at Chalmers University of Technology, Gothenburg, Sweden. His current research interests include integrated sensing and communications, RIS-aided localization and sensing, and hardware impairments in beyond 5G/6G systems.
\end{IEEEbiographynophoto}

\vspace{-0.5cm}
\begin{IEEEbiographynophoto}{Adham Sakhnini} (adham.sakhnini@imec.be) is a PhD student at IMEC and KU Leuven. His research interests revolve around the analysis and design of radar and communication systems.
\end{IEEEbiographynophoto}

\vspace{-0.5cm}
\begin{IEEEbiographynophoto}{Nicol\`o Decarli} (nicolo.decarli@cnr.it) is a researcher at the National Research Council (IEIIT), and an affiliate of WiLAB-CNIT, Italy.
His research interests include next-generation radio communication, localization, and sensing.
\end{IEEEbiographynophoto}

\vspace{-0.5cm}
\begin{IEEEbiographynophoto}{Sofie Pollin} (sofie.pollin@esat.kuleuven.be) is a full professor with the Electrical Engineering Department, KU Leuven. Her research interests include networked systems that require networks that are ever more dense, heterogeneous, battery-powered, and spectrum-constrained.
\end{IEEEbiographynophoto}

\vspace{-0.5cm}
\begin{IEEEbiographynophoto}{Davide Dardari} (davide.dardari@unibo.it)  is a professor at the University of Bologna, I-40136 Bologna, Italy, and an affiliate of WiLAB-CNIT, Italy. His research interests include radio localization, next-generation wireless communications, and smart radio environments. 
\end{IEEEbiographynophoto}

\vspace{-0.5cm}
\begin{IEEEbiographynophoto}{Henk Wymeersch} (henkw@chalmers.se)
is a professor at Chalmers University
of Technology, 41296 Gothenburg, Sweden. His research interests include 5G
and beyond 5G radio localization and
sensing.
\end{IEEEbiographynophoto}

\vfill

\end{document}